\documentclass[twocolumn,longbibliography,preprintnumbers,amsmath,amssymb,10pt]{revtex4-2}
\usepackage{iftex}
\ifTUTeX
\usepackage{fontspec}
\else
\usepackage[T1]{fontenc}
\usepackage[utf8]{inputenc} 
\DeclareUnicodeCharacter{200B}{{\hskip 0pt}}
\fi
\usepackage[LGR,T1]{fontenc}
\usepackage{textgreek}
\usepackage{natbib}
\usepackage[english]{babel}
\usepackage[utf8]{inputenc}

\usepackage{enumitem}
\usepackage{braket}
\usepackage{footnote}
\usepackage{tablefootnote}
\usepackage[margin=1in]{geometry}
\usepackage{amsmath,amssymb,amsfonts,mathrsfs}
\usepackage{booktabs}
\usepackage{mathtools}
\usepackage{bm}
\usepackage{dcolumn}
\usepackage{graphicx}   
\usepackage{latexsym}
\usepackage{cmbright}
\usepackage[OT1]{fontenc}
\usepackage{kpfonts}
\usepackage[acronym]{glossaries}
\usepackage{physics}
\usepackage{siunitx}
\usepackage{comment}
\DeclareSIUnit{\persqrthz}{\ensuremath{\text{Hz}^{-1/2}}}
\usepackage[svgnames]{xcolor}
\usepackage[final]{hyperref}
\hypersetup{%
	colorlinks=true,
	citecolor=Blue,
}%

\newcommand{\myhyperref}[1]{\hyperref[#1]{\ref{#1}}}

\begin{document}
	
	\title{A class of rotating metrics in the presence of a scalar field}
	
	\author{Behrouz Mirza}
	\email{b.mirza@iut.ac.ir}
	
	\author{Parichehr Kangazian Kangazi}
	\email{p.kangazian@ph.iut.ac.ir}
	
	\author{Fatemeh Sadeghi}
	\email{fatemeh.sadeghi96@ph.iut.ac.ir}
	\affiliation{Department of Physics, Isfahan University of Technology, Isfahan 84156-83111, Iran}

	\begin{abstract}
		\vspace{5mm}
		\vspace{5mm}
		\begin{center}
			\textbf{Abstract}
		\end{center}
		We consider a class of three parameter static and axially symmetric metrics that reduce to the Janis-Newman-Winicour (JNW) and $ \gamma$-metrics in certain limits of the parameters. We obtain rotating form of the metrics that are asymptotically flat, stationary and axisymmetric. In certain values of the parameters, the solutions represent the rotating JNW metric, rotating $ \gamma$-metric and Bogush-Gal’tsov (BG) metric. The singularities of rotating metrics are investigated. Using the light-ring method, we obtain the quasi normal modes (QNMs) related to rotating metrics in the eikonal limit. Finally, we investigate the precession frequency of a test gyroscope in the presence of the rotating metrics.
	\end{abstract}
	
	
	\newacronym{e2e}{E2E}{End-To-End}
	\newacronym{inrep}{INREP}{Initial Noise REduction Pipeline}
	\newacronym{tdi}{TDI}{Time Delay Interferometry}
	\newacronym{ttl}{TTL}{Tilt-To-Length couplings}
	\newacronym{dfacs}{DFACS}{Drag-Free and Attitude Control System}
	\newacronym{ldc}{LDC}{LISA Data Challenge}
	\newacronym{lisa}{LISA}{the Laser Interferometer Space Antenna}
	\newacronym{emri}{EMRI}{Extreme Mass Ratio Inspiral}
	\newacronym{ifo}{IFO}{Interferometry System}
	\newacronym{grs}{GRS}{Gravitational Reference Sensor}
	\newacronym{tmdws}{TM-DWS}{Test-Mass Differential Wavefront Sensing}
	\newacronym{ldws}{LDWS}{Long-arm Differential Wavefront Sensing}
	\newacronym[	plural={MOSAs},
	first={Moving Optical Sub-Assembly},
	firstplural={Moving Optical Sub-Assemblies}
	]{mosa}{MOSA}{Moving Optical Sub-Assembly}
	\newacronym{siso}{SISO}{Single-Input Single-Output}
	\newacronym{mimo}{MIMO}{Multiple-Input Multiple-Output}
	\newacronym[plural=MBHB's, firstplural=Massive Black Holes Binaries (MBHB's)]{mbhb}{MBHB}{Massive Black Holes Binary}
	\newacronym{cmb}{CMB}{Cosmic Microwave Background}
	\newacronym{sgwb}{SGWB}{Stochastic Gravitational Waves Background}
	\newacronym{pta}{PTA}{Pulsar Timing Arrays}
	\newacronym{gw}{GW}{Gravitational Wave}
	\newacronym{snr}{SNR}{Signal-to-Noise Ratio}
	\newacronym{pbh}{PBH}{Primordial Black Holes}
	\newacronym{psd}{PSD}{Power Spectral Density}
	\newacronym{tcb}{TCB}{Barycentric Coordinate Time}
	\newacronym{bcrs}{BCRS}{Barycentric Celestial Reference System}
	\newacronym{lhs}{LHS}{Left-Hand Side}
	\newacronym{rhs}{RHS}{Right-Hand Side}
	\newacronym{mcmc}{MCMC}{Monte-Carlo Markov Chains}
	\newacronym{cs}{CS}{Cosmic Strings}
	\newacronym{ssb}{SSB}{Solar System Barycentric}
	\newacronym{oms}{OMS}{Optical Metrology System}
	\newacronym{dof}{DoF}{Degree of Freedom}
	\newacronym{eob}{EOB}{Effective One-Body}
	\newacronym{pn}{PN}{Post-Newtonian}
	\newacronym{cce}{CCE}{Cauchy-Characteristic Evolution}
	\newacronym{imr}{IMR}{Inspiral-Merger-Ringdown}
	\newacronym{scird}{SciRD}{LISA Science Requirement Document}
	
	
	%
	\maketitle
	
	\section{\label{Intro}Introduction}
In general relativity, singularities appear at those points where the curvature invariants of space-time become infinite. In the 1960s, it was shown that the existence of the singularity is an absolute consequence of the theory of general relativity
~\cite{penrose1965gravitational, hawking1967occurrence, hawking1970singularities}.  
Such singularities also exist in the form of infinite mater density.
They are classified into two types including black holes and naked singularities. Unlike black holes, which singularities are hidden by an event horizon and therefore cannot be directly observed, naked singularities may show other types of effects from outside.

The strong cosmic censorship conjecture asserts that no naked singularities can exist and they are not visible to any observer \cite{4, 5}. However, some recent theoretical investigations present us with some striking physical processes that create naked singularities
\cite{105, 106, 6}. 
Black holes can have only charge, mass and angular momentum based on the no-hair theorem. However, other astrophysical objects such as planets, neutron stars, white dwarfs etc can have higher-order multiple moments and, therefore, both of these objects and naked singularities that have the same multiple moments have similar exterior metric. 

Two of the solutions of Einstein equations that represent naked singularity are the $ \gamma$-metric (also known as $ q$-metric or Zipoy-Voorhees metric)
\cite{darmois1927memorial, erez1959gravitational, zipoy1966topology, voorhees1970static}
and the Janis-Newman-Winicour (JNW) metric
\cite{fisher1999scalar, janis1968reality, wyman1981static}, which are asymptotically flat and axially symmetric metrics. The $ \gamma$-metric is written in vacuum, while the singularity of the JNW metric is essentially sourced by existence of a massless scalar field. The JNW metric reduces to the Schwarzschild metric in the absence of the scalar field. Until today, many studies have been done on $ \gamma$-metric and JNW space-time. For example see the following papers and references therein.
Geodesics of JNW metric was studied in 
\cite{chowdhury2012circular, turimov2018axially}. 
The role of a scalar field in gravitational lensing studied in
\cite{virbhadra1998role, dey2008gravitational} 
and high energy collisions in this space-time was investigated in 
\cite{patil2012acceleration}. 
For study in $ \gamma$-metric see
\cite{herrera2000geodesics, richterek2002einstein, chakrabarty2018unattainable, abdikamalov2019black, toshmatov2019harmonic, benavides2019charged, 1812, allahyari2020quasinormal, chakrabarty2022effects, hajibarat2022gamma, li2022constraining,  chakrabarty2023constraining} and references.  

In this article, we study a general three parameter metric in the presence of the scalar field, from which JNW and $ \gamma$-metrics can be obtained in certain values of the parameters. Considering that almost most of celestial bodies rotate around themselves, it is interesting to derive a rotating form of this class of three parameter metrics. It should be noted that, despite many studies, rotating solutions for the JNW metric that have been proposed to date, were not correct, because, the metrics don't satisfy the Einstein field equations 
\cite{pirogov2013towards, bogush2020generation}. In this paper, we introduce a class of stationary and axially symmetric rotating metrics that include the rotating JNW, Bogush-Gal’tsov (BG)
\cite{bogush2020generation} and rotating $ \gamma$-metric and also many more new rotating metrics that can be obtained in certain limits of the three  parameters.

One way to better understand black holes and naked singularities is to study their interactions with the surrounding environment because it is impossible to find them in isolation. These interactions usually result in space-time perturbations. For the first time, Regge and Wheeler investigated a category of gravitational perturbations  in Schwarzschild geometry, even before the concept of a black hole was born
\cite{regge1957stability}.
In 1970s, numerical analysis of these perturbations  showed that when a compact astrophysical object is perturbed, gravitational  waves with damped oscillations are emitted. These waves are known as quasi-normal modes (QNMs).

When space-time is perturbed, it emits gravitational waves that change over time, so that initially we have a burst of radiation in a short time, then over a long period of time, the black hole loses a lot of energy by emitting QNMs. Finally, after a very long period of time, the QNMs are suppressed.

Analytical and numerical analysis of QNMs shows that the frequencies obtained for QNMs depend only on the parameters of the astrophysical objects, while the amplitude of these QNMs depends on the perturbation source of the oscillation.  There are several methods to calculate QNMs, for reviews see
\cite{nakamura1987general, kokkotas1999quasi, nollert1999quasinormal, ferrari2008quasi, berti2009quasinormal, konoplya2011quasinormal}
and references therein. 

We will study QNMs of the rotating metrics in the following sections. We also investigate the precession frequency due to the Lense-Thirring(LT) and de Sitter effects of a gyroscope in the case of rotating space-times. A test gyroscope is a suitable device for studying the properties of rotating space times.

The paper is organized as follows, in Sec. \ref{2}, we consider a class of  three parameter metrics that at certain values of the parameters represent the JNW and $ \gamma$-metrics. Then, in Sec. \ref{3}, we generalize the static metrics and obtain a  class of axially symmetric rotating solutions of Einstein's equations and explain special values of the parameters that one may derive rotating $ \gamma$-metric and rotating JNW solution. In Sec. \ref{4}, we numerically investigate the singularities of rotating metrics. In Sec. \ref{5}, the QNMs of the axially symmetric rotating metrics are derived in the eikonal limit. In Sec. \ref{6100}, we will consider a test gyroscope and calculate the precession frequency of it in the rotating space-times that we introduced. Finally, Sec. \ref{7100} is devoted to the conclusions.

In this paper $ c = G = \hbar = 1 $.
\section{A class of three parameter static metrics in the presence of a radial scalar field}\label{2}
In this section, we are going to represent a class of three parameter  metrics that are static and axially symmetric and reduce to the Janis-Newman-Winicour (JNW) and $ \gamma$-metrics in certain limits of the parameters. At the beginning, we write the Einstein-Hilbert action when minimally coupled to a massless scalar field as follows
\begin{equation}\label{1_label}
	S = \frac{1}{16 \, \pi} \int d^4 x \, \sqrt{-g} \, [R - 8 \, \pi \, g^{\mu \nu} \; \partial_{\mu} \varphi \, \partial_{\nu} \varphi].
\end{equation}
The field equations related to the above action are
\begin{equation}\label{2_label}
	R_{\mu \nu} = 8 \, \pi \, (\partial_{\mu} \varphi) \, (\partial_{\nu} \varphi), \qquad \Box \varphi = 0.
\end{equation}
We can write a general solution of the equations of motion by the following class of three parameter static metrics
\cite{Azizallahi2023}
\begin{equation}\label{3_label}
	d s^2 = - f^\gamma \, dt^2 + f^\mu \, k^\nu \, \big( \frac{d r^2}{f} + r^2 \, d \theta^2 \big) + r^2 f^{1 - \gamma} \sin^2 \theta \, d \phi^2,
\end{equation}
where,
\begin{equation}\label{4_label}
	f(r) = 1 - \frac{2 \, m}{r}, \qquad k(r, \theta) = 1 - \frac{2 \, m}{r} + \frac{m^2 \sin^2 (\theta)}{r^2},
\end{equation}
and $ \mu $ and $ \nu $ are defined as 
\begin{equation}\label{5_label}
	\mu + \nu = 1 - \gamma.
\end{equation}
where the physical mass is given by $ M = \gamma \, m $. A solution for the scalar field can be obtained as follows 
\begin{equation}\label{6_label}
	\varphi(r) = \sqrt{\frac{1 - \gamma^2 - \nu}{16 \, \pi}} \; \ln \big( 1 - \frac{2 \, m}{r} \big).
\end{equation}
According to Eqs. \eqref{5_label} and \eqref{6_label}, the following conditions must be satisfied
\begin{equation}\label{201_label}
	\nu \leq 1 - \gamma^2, \qquad \mu \geq \gamma^2 - \gamma.
\end{equation}
Furthermore, all components of the Ricci tensor are zero except the $ R_{r r} $, which is $ R_{r r} = 8 \, \pi \, \partial_r \varphi \, \partial_r \varphi $.

Different limits of metric in Eq. \eqref{3_label} leads to the following well known metrics. Let us suppose that $ \mu = 1 - \gamma \, (\nu = 0) $, therefore, the metric \eqref{3_label} reduces to the JNW metric. Also, using the following values for the parameters
\begin{equation}\label{7_label}
	\nu = 1 - \gamma^2, \qquad \mu = \gamma^2 - \gamma.
\end{equation} 
the metric in Eq. \eqref{3_label} becomes $ \gamma$-metric and the scalar field in Eq. \eqref{6_label} vanishes. It should be noted that vacuums are different in general relativity in the presence of a scalar field and without the scalar field.

For other values of parameters $ \mu $ and $ \nu $, metric in Eq. \eqref{3_label} interpolate between JNW and $ \gamma$-metric and represent infinitely many different metrics  with naked singularities. Two interesting cases of Eq. \eqref{3_label} are $ \mu = - \nu $, $ \gamma = 1 $ and $ \mu=0, \nu=1-\gamma $, that we will study their rotating forms  in the following sections.

\section{A class of rotating naked singularities}\label{3}
There are different ways to obtain rotating solutions of the Einstein's equations. For example the Janis-Newman algorithm can be used to obtain some type of rotating metrics 
\cite{newman1965note, newman1965metric, kamenshchik2023newman}. 
By using Ernst's method, it is possible to obtain the Kerr-Newman solution  and  rotating $ \gamma$-metric
\cite{ernst1968new1, ernst1968new, kinnersley1977symmetries1, kinnersley1977symmetries, kinnersley1978symmetries1, kinnersley1978symmetries, hoenselaers1979symmetries1, hoenselaers1979symmetries}.

In this section, we first introduce rotating form of the general class of metrics in Eq. \eqref{3_label} and then explain about the related Ernst potential.

Consider the following stationary and axisymmetric line element which is written in the prolate coordinates $ (t, x, y, \phi) $
\begin{equation}\label{8_label}
	\begin{aligned}
		d s^2 = & - f \, (dt - \omega \, d \phi)^2 + \frac{\sigma^2}{f} \, \big[ e^{2 \, \eta} \, (x^2 - y^2)\\
		& \times ( \frac{d x^2}{x^2 - 1} + \frac{d y^2}{1 - y^2} ) + (x^2 - 1) \, (1 - y^2) \, d \phi^2 \big].
	\end{aligned}
\end{equation}
where, $ \sigma $ is a positive constant and $ f $, $ \eta $, and $ \omega $ depond only on x and y. The rotating metric functions are
\begin{equation}\label{9_label}
	f = \frac{A}{B},
\end{equation} 
\begin{equation}\label{10_label}
	\omega = -2 \, \big( a + \sigma \, ( \frac{C}{A} ) \big),
\end{equation} 
\begin{equation}\label{11_label}
	e^{2 \eta} = \frac{1}{4} \, \big( 1 + \frac{m}{\sigma} \big)^2 \; \frac{A}{(x^2 - 1)^{1 + q}} \; \big( \frac{x^2 - 1}{x^2 - y^2} \big)^{1 - \nu}.
\end{equation}
Here $ a $ is the rotation parameter and $ q = \gamma - 1 $ is the quadrupole parameter and $ \nu $ was defined in \eqref{5_label}. Moreover
\begin{equation}\label{12_label}
	\begin{aligned}
		A & = \, a_+ \, a_- + b_+ \, b_-,\\
		B & = \, a_+^2 + b_+^2,\\
		C & = \, (x + 1)^q \, \big[ x \, (1 - y^2) \, (\lambda + \xi) \, a_+\\
		& + y \, (x^2 - 1) \, (1 - \lambda \, \xi) \, b_+ \big],
	\end{aligned}
\end{equation}
and
\begin{equation}\label{13_label}
	\begin{aligned}
		& a_\pm = (x \pm 1)^q \; \big[ x \, (1 - \lambda \, \xi) \pm (1 + \lambda \, \xi) \big],\\
		& b_\pm = (x \pm 1)^q \; \big[ y \, (\lambda + \xi) \mp (\lambda - \xi) \big],\\
		& \lambda = \alpha \, \big( x^2 - 1 \big)^{-q} \; (x + y)^{2 \, q},\\
		& \xi = \alpha \, \big( x^2 - 1 \big)^{-q} \; (x - y)^{2 \, q},\\
		& \alpha = - \frac{a}{m + \sigma},\\
		&\sigma = \sqrt{m^2 - a^2}.
	\end{aligned}
\end{equation}
Besides, relation to the spherical coordinates are as follows
\begin{equation}\label{14_label}
	x = \frac{r - m}{\sqrt{m^2 - a^2}}, \qquad y = cos \theta.
\end{equation}
The solution for scalar field equation ($ \Box \varphi = 0 $) is given by 
\begin{equation}\label{15_label}
	\varphi(r) = \sqrt{\frac{1 - \gamma^2 - \nu}{16 \, \pi}} \; \ln \big( \frac{r - m - \sqrt{m^2 - a^2}}{r - m + \sqrt{m^2 - a^2}} \big).
\end{equation}
The Ricci tensor related to the metric in Eq. \eqref{8_label} can be obtained as follows
\begin{equation}\label{16_label}
	R_{\mu \nu} = 8 \pi 
	\begin{pmatrix}
		0 & 0 & 0 & 0 \\ 0 & \partial_r \varphi \, \partial_r \varphi & 0 & 0 \\ 0 & 0 & 0 & 0 \\ 0 & 0 & 0 & 0
	\end{pmatrix}.
\end{equation}
One can check the correctness of this class of rotating solutions simply by using Mathematica or Maple packages and a laptop.  In certain values of the parameters the metric that defined in \eqref{8_label} represents rotating form of the following metrics:

a) By choosing $ \nu =  1 - \gamma^2 $ in equation \eqref{11_label}, we obtain the rotating $ \gamma$-metric
\cite{toktarbay2014stationary, frutos2018relativistic}.

b) Assuming a non-zero value for $ \nu $ in equation \eqref{11_label} and setting $ q $ equal to zero ($ \gamma = 1 $) in all equations, we will have Bogush-Gal’tsov (BG) metric
\cite{bogush2020generation}
\begin{equation}\label{8xa_label}
	\begin{aligned}
		ds^2 & = -\frac{\Delta}{\rho^2} \, (dt - a \sin^2 \theta \, d\phi)^2\\
		& + \frac{\rho^2}{\Delta} \, \big[ 1+ \frac{(m^2 - a^2) \, \sin^2 \theta}{\Delta} \big]^\nu \; dr^2 \\
		& + \rho^2 \, \big[ 1+ \frac{(m^2 - a^2) \, \sin^2 \theta}{\Delta} \big]^\nu \; d\theta^2\\
		& + \frac{\sin^2 \theta}{\rho^2} \; [a\, dt - (r^2 + a^2) \, d \phi]^2, \\
	\end{aligned}
\end{equation}
where $ \Delta $ and $ \rho $ are as follows
\begin{equation}\label{8xa_label}
	\begin{aligned}
		& \Delta = r^2 + a^2 - 2 \, m \, r, \\
		& \rho = \sqrt{r^2 + a^2 \cos^2 \theta}. \\
	\end{aligned}
\end{equation}
Considering Eq. \eqref{15_label} and the fact that $ \gamma =1 $, then we must always have $ \nu \le 0 $.

The Ricci scalar for the BG metric is as follows
\begin{equation}\label{riccibg_label}
	R= \frac{2\,\nu\,(a^2-m^2)\,\big[a^2+r\,(r-2\,m)+(m^2-a^2)\,\sin^2\theta\big]^{-\nu}}{\big[a^2+r\,(r-2\,m)\big]^{1-\nu}\,(r^2+a^2\,\cos^2\theta)}.
\end{equation}
The Ricci scalar becomes infinite at $ r=m\pm \sqrt{m^2-a^2} $. Also, the Ricci and Kretschmann scalars are curvature invariants and their singularities coincide.

c) Another interesting limit is $ \mu = 0 $ and $ \nu = 1 - \gamma $, that leads to a new form of metric that we will study its QNMs in Sec. \ref{5}. In this metric ($ \mu=0,\,\nu=1-\gamma $), according to the Eq. \eqref{15_label}, we have
\begin{equation}\label{mgam_label}
	\gamma-\gamma^2\ge 0,\quad\Rightarrow \quad 0\le \gamma \le 1.
\end{equation}
d) Finally, if we set $ \nu $ equal to zero and for non-zero values of $ q $, we obtain for the first time a correct form of the rotating JNW metric. In this condition if we put $ a = 0 $, the metric reduces to the JNW metric. For non-zero values of $ a $ and $ q = 0 $ ($ \gamma = 1 $), the rotating metric becomes the Kerr metric. In the rotating JNW metric ($ \nu=0,\,\mu=1-\gamma $), according to Eq. \eqref{15_label}, we have
\begin{equation}\label{mnu_label}
	1-\gamma^2\ge 0, \quad \Rightarrow \quad - 1<\gamma<1.
\end{equation}

Now, we describe a method for obtaining metric functions. For this purpose, we write the metric in Eq. \eqref{9_label} in $ (t, \rho, z, \phi) $ coordinates
\begin{equation}\label{500_label}
	ds^2 = - f \, (d t - \omega \, d \phi)^2 + \frac{\sigma^2}{f} \, \big[ e^{2 \, \eta} \, (d \rho^2 + d z^2) + \rho^2 \, d \phi^2 \big],
\end{equation}
where
\begin{equation}\label{500a_label}
	\begin{aligned}
		& t = t, \quad \phi = \phi, \quad \rho = (x^2 - 1)^{\frac{1}{2}} \; (1 - y^2)^{\frac{1}{2}}, \quad z = x \, y,\\
		& x = \frac{1}{2} \, (R_+ + R_-), \quad y = \frac{1}{2} \, (R_+ - R_-),\\
		& R_\pm = \sqrt{\rho^2 + (z \pm 1)^2}.
	\end{aligned}
\end{equation}
In the following we are going to obtain metric functions.

Using $ R_{t t} $ and $ R_{t \phi} $, we can derive the following equations (To see the components of the Ricci tensor, refer to the appendix)
\begin{equation}\label{502_label}
	\begin{aligned}
		& f \, (\partial_\rho^2 \, f + \partial_z^2 \, f + \frac{\partial_\rho \, f}{\rho}) - (\partial_\rho \, f)^2 - (\partial_z \, f)^2\\
		& + \frac{f^4}{\rho^2 \, \sigma^2} \; \big[ (\partial_\rho \, \omega)^2 + (\partial_z \, \omega)^2 \big]  = 0,
	\end{aligned}
\end{equation}
\begin{equation}\label{504_label}
	\partial_\rho \, (\frac{f^2 \, \partial_\rho \, \omega}{\rho}) + \partial_z  \, (\frac{f^2 \, \partial_z \, \omega}{\rho}) = 0.
\end{equation}
According to Eq. \eqref{504_label}, a function $ u $ can be defined as follows
\begin{equation}\label{505_label}
	\partial_\rho \, u = - \frac{f^2 \, \partial_z \, \omega}{\rho \, \sigma}, \qquad \partial_z \, u =  \frac{f^2 \, \partial_\rho \, \omega}{\rho \, \sigma}.
\end{equation}
By using Eq. \eqref{505_label}, we can write Eq. \eqref{502_label} as follows
\begin{equation}\label{510_label}
	f \, \nabla^2 \, f - (\partial_\rho \, f)^2 - (\partial_z \, f)^2 + (\partial_\rho \, u)^2 +  (\partial_z \, u)^2 = 0,
\end{equation}
that  $ \nabla^2 f  $ is equal to
\begin{equation}\label{511_label}
	\nabla^2 \, f  = \partial_\rho^2 \, f + \partial_z^2 \, f + \frac{\partial_\rho \, f}{\rho}.
\end{equation}
Also, according to Eq. \eqref{505_label}, we can write the following relation for $ u $
\begin{equation}\label{512_label}
	f \, \nabla^2 \, u = 2 \, (\partial_\rho \, f \, \partial_\rho \, u + \partial_z \, f \, \partial_z \, u).
\end{equation}
Using the $ R_{\rho z} $ and Eq. \eqref{505_label}, we have
\begin{equation}\label{506_label}
	\partial_z \, \eta = \frac{\rho}{2} \, \big[ \frac{1}{f^2} \, (\partial_\rho \, f \, \partial_z \, f + \partial_\rho \, u \, \partial_z \, u) + 16 \, \pi \, \partial_\rho \, \varphi \, \partial_z \, \varphi \big].
\end{equation}
Also, using $ R_{\rho \rho} $ and $ R_{z z} $ and Eq. \eqref{505_label}, we have
\begin{equation}\label{507_label}
	\begin{aligned}
		\partial_\rho \, \eta & = \frac{\rho}{4} \, \big\{ \frac{1}{f^2} \, \big[ (\partial_\rho \, f )^2 - (\partial_z \, f)^2 + (\partial_\rho \, u)^2 - (\partial_z \, u)^2 \big]\\
		& + 16 \, \pi \, [(\partial_\rho \, \varphi)^2 - (\partial_z \, \varphi)^2] \big\}.
	\end{aligned}
\end{equation}
Now by using prolate coordinates $ (t, x, y, \phi) $, Eqs. (\ref{505_label}, \ref{506_label} and \ref{507_label}) can be represented as follows
\begin{equation}\label{24_label}
	\partial_x \, \omega = \sigma \, (1 - y^2) \, f^{-2} \; \partial_y u, \quad \partial_y \, \omega = \sigma \, (1 - x^2) \, f^{-2} \; \partial_x u,
\end{equation}
\begin{equation}\label{25_label}
	\begin{aligned}
		\partial_x \, \eta & = \frac{(1 - y^2) \, f^{- 2}}{4 \, (x^2 - y^2)} \, \big\{ x \, (x^2 - 1) \, \big[ (\partial_x \, f)^2 + (\partial_x \, u)^2 \big]\\ & + x \, (y^2 - 1) \, \big[ (\partial_y \, f)^2 + (\partial_y \, u)^2 \big]\\
		& - 2 \, y \, (x^2 - 1) \, (\partial_x \, f \, \partial_y \, f + \partial_x \, u \, \partial_y \, u)  \big\}\\
		& + 4 \, \pi \, \frac{(1 - y^2)}{(x^2 - y^2)} \, x \, (x^2 - 1) \, (\partial_x \, \varphi)^2,
	\end{aligned}
\end{equation}
\begin{equation}\label{26_label}
	\begin{aligned}
		\partial_y \, \eta = & \frac{(x^2 - 1) \, f^{- 2}}{4 \, (x^2 - y^2)} \, \big\{ y \, (x^2 - 1) \, \big[ (\partial_x \, f)^2 + (\partial_x \, u)^2 \big]\\
		& - y \, (1 - y^2) \, \big[ (\partial_y \, f)^2 + (\partial_y \, u)^2 \big]\\
		& + 2 \, x \, (1 - y^2) \, (\partial_x \, f \, \partial_y \, f + \partial_x \, u \, \partial_y \, u)  \big\}\\
		& +  4 \, \pi \, \frac{(x^2 - 1)^2 \, y}{(x^2 - y^2)} \, (\partial_x \, \varphi)^2.
	\end{aligned}
\end{equation}
Now we define a complex function $ E $ as follows
\begin{equation}\label{18_label}
	E = f + i \, u.
\end{equation}
Now, with the definition \eqref{18_label}, we can display Eqs. \eqref{510_label} and \eqref{512_label} in prolate coordinates with the following relation
\begin{equation}\label{17_label}
	(E + E^*) \, \Delta E = 2 \, (\nabla E)^2,
\end{equation}
where $ E^* $ is the complex conjugate of \eqref{18_label} and we define
\begin{equation}\label{19_label}
	\begin{aligned}
		\Delta \equiv & \,  \sigma^{- 2} \, (x^2 - y^2)^{- 1} \; {\partial_x \, \big[ (x^2 - 1) \, \partial_x \big] + \partial_x \, \big[ (x^2 - 1) \, \partial_x \big]},\\
		\nabla \equiv & \,  \sigma^{- 1} \, (x^2 - y^2)^{- \frac{1}{2}} \; {\hat i \, \big[ (x^2 - 1)^{\frac{1}{2}} \, \partial_x \big]  + \hat j \, \big[ (1 - y^2)^{\frac{1}{2}} \, \partial_x \big]}.
	\end{aligned}
\end{equation}
It can be shown that the following Ernst potential consistently explains the rotating solutions
\begin{equation}\label{20_label}
	E = \big( \frac{x - 1}{x + 1} \big)^q \; \frac{x - 1 + (x^2 - 1)^{- q} \; d_+}{x + 1 + (x^2 - 1)^{- q} \; d_-},
\end{equation}
with
\begin{equation}\label{21_label}
	\begin{aligned}
		d_\pm & = - \alpha^2 \, (x \pm 1) \, h_+  \, h_- (x^2 - 1)^{- q}\\
		& + i \, \alpha \, \big[ y \, (h_+ + h_-) \pm (h_+ - h_-) \big],
	\end{aligned}
\end{equation}
and
\begin{equation}\label{22_label}
	h_\pm = (x \pm y)^{2 \, q}.
\end{equation}
Now, if we take the derivative of Eq. \eqref{25_label} with respect to $ y $, the obtained answer must be equal to the derivative of Eq. \eqref{26_label} with respect to $ x $. This condition gives us a second order differential equation that we can use to obtain the scalar field as follows 
\begin{equation}\label{27_label}
	\varphi^{\prime \prime} + \frac{2 \, x}{x^2 - 1} \, \varphi^{\prime} = 0, \quad \Rightarrow \quad \varphi = c_1 \, \ln \big( \frac{x - 1}{x + 1} \big) + c_2.
\end{equation}
where, constant numbers can be obtained from Einstein's equations and boundary conditions. Therefore, using this conditions the above constants $ c_1 $ and $ c_2 $ can be written as follows
\begin{equation}\label{28_label}
	c_1 = \sqrt{\frac{1 - \gamma^2 - \nu}{16 \, \pi}}, \qquad c_2 = 0.
\end{equation}

Therefore we have obtained Eq. \eqref{15_label} for the scalar field. It should be noted that using the Janis-Newman algorithm does not give us a correct form of rotating JNW metric. However, our solution is exact and has a  correct form for a large class of rotating metrics that include correct form of the JNW metric. 
\section{Kretschmann scalar for rotating metrics}\label{4}
\begin{figure}[t]
\centering
\includegraphics[width=1.0\columnwidth, trim={0.2cm 0.0cm 0.2cm 0.0cm}, clip]{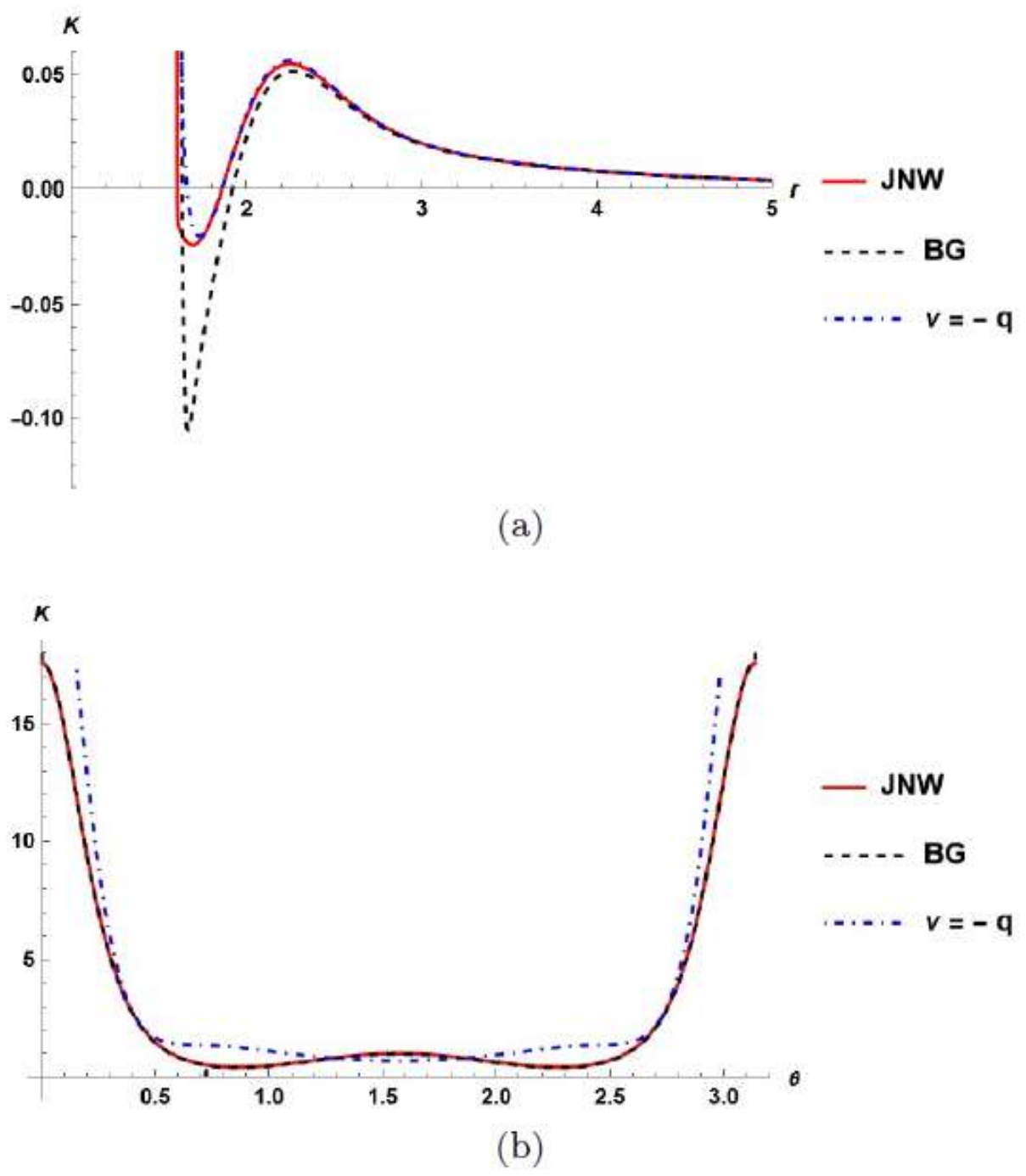}
\caption{The Kretschmann scalar $ K(r , \theta) $ for $ m = 1 $ and $ a = 0.8 $. Panel (a): $ K(r , \pi/2) $, $ q = -0.015 $ for rotating JNW metric (solid red), $ \nu = - 0.08 $ for BG metric (black dash) and  $ q = - 0.015 $ for rotating metric at $ \nu=-q $ (blue dot-dash). K diverges at near $ r = 1.6 $ and for all values of $ \theta $, the singularities exist and occur irrespectively of the value of $\theta$. Panel (b): $ K(18 , \theta) $, where $ q = -0.015 $ for rotating JNW metric (solid red), $ \nu = - 0.08 $ for BG metric (black dash) and  $ q = - 0.35 $ for rotating metric at $ \nu=-q $ (blue dot-dash). K is a symmetric function with respect to $ \theta = \pi / 2 $.}
\label{1fig}
\end{figure}

In Sec. \ref{3}, we obtained the  rotating metrics, which are axially symmetric and asymptotically flat. In this section, we shall study the singularity structure of these metrics. For this purpose, we investigate the Kretschmann scalar ($ K(r , \theta) $) related to these metrics. Although it is possible to calculate the Kretschmann scalar accurately, the answer is long and complicated. Therefore, we will investigate it numerically. We expect that our results will be such that, firstly, the Kretschmann scalar tends to zero at infinity, secondly, its values at constant $ r $ are symmetric with respect to $ \theta = \pi/2 $ and thirdly, the radius of the outer singularity is $ r_s = m + \sqrt{m^2 - a^2} $, and as a result, the graphs around this value diverge at a constant $ \theta $. The Kretschmann scalar is depicted in Fig. \ref{1fig}.

According to Fig. \ref{1fig}, it can be clearly seen that the behavior of the Kretschmann scalar fulfill all our expectations.
\section{QNMs in the eikonal limit and for slowly rotating metrics}\label{5}
In this section, we are going to calculate the QNMs of some special cases of the rotating metrics. For deriving QNMs, we use the light-ring method.  More explanation about this method is given in Refs.
\cite{ferrari1984oscillations, ferrari1984new, mashhoon1985stability}.

We shall study rotating JNW metric ($ \nu = 0 $ or $ \mu = 1 - \gamma $), BG metric ($ \mu = - \nu $ 0r $ \gamma = 1 $) and the rotating metric at $ \nu=-q $ as follows.
\subsection{QNMs of rotating JNW metric $ (\nu=0) $}\label{62}
First, we investigate QNMs related to the rotating JNW metric. We replace transformations \eqref{14_label} in the metric \eqref{8_label}. Then, by putting $ \nu = 0 $ in Eq. \eqref{11_label} and expanding the result to second order in $ a $ and first order in $ q $ and, for simplicity, also assuming that $ q $ is of order $ a^2 $ and therefore terms proportional to $ a\,q $ are negligible, we have
\begin{equation}\label{50_label}
\begin{aligned}
d s^2 & = - \, \big\{ 1 - \frac{2 \, m}{r}  + \frac{2 \, m \, a^2 \cos^2\theta}{r^3}\\
& + q \, (1 - \frac{2 \, m}{r}) \, \ln \small( 1 - \frac{2 \, m}{r} \small) \big\} \, d t^2\\
& - \frac{1}{r^3 \, (1 - 2 \, m/r)^2} \, \big\{ a^2 \, (r \, \sin^2 \theta + 2 \, m \, \cos^2 \theta)\\
& + r^3\,(1-\frac{2\,m}{r})\,\big[q\,\ln\small(1-\frac{2\,m}{r}\small)-1\big]\big\}\, d r^2\\
& + \big\{ r^2 + a^2 \, \cos^2 \theta - q \, r^2 \, \ln \small( 1 - \frac{2 \, m}{r} \small) \big\} \, d \theta^2\\
& + \sin^2 \theta \, \big\{ r^2 +  \frac{a^2}{r} \, (r + 2 \, m \, \sin^2 \theta)\\
& - q \, r^2 \, \ln \small( 1 - \frac{2 \, m}{r} \small) \big\} \, d \phi^2\\
&+ \frac{4 \, a \, m}{r} \sin^2 \theta \, d \phi \, d t.
\end{aligned}
\end{equation}
It should be noted that we will use these approximations in all subsequent calculations.

If we consider QNMs as $ Q = \Omega + i \, \Gamma $, in the eikonal limit by using the light-ring method $ \Omega $ can be written as $ \pm \,  j \, \Omega^\prime $, where $ j \in \mathbb{Z} $ and $ \Omega^\prime $ is the orbital frequency of the null rays on circular orbits. In our axisymmetric space-time background the massless wave perturbations can be written as superposition of the following eigenmodes 
\begin{equation}\label{51_label}
	e^{i \, (\Omega \, t - \iota \, \phi)} \, S_{\Omega \, j \, \iota \, s} (r, \theta).
\end{equation}
where, $ \Omega $ and $ S $ are the wave's frequency and spin respectively. Also, $ j $ and $ \iota $ are angular momentum parameters with condition $ \lvert \iota \rvert \le j $. In the eikonal limit, it is assumed that $ \Omega \gg 1/M $ and $ \lvert \iota \rvert = j \gg 1 $. It can be seen from the results of numerical calculations for QNMs of black holes such as 
\cite{ferrari1984oscillations, ferrari1984new, mashhoon1985stability}
that the oscillation frequency of such perturbations is as follows
\begin{equation}\label{53_label}
	\Omega = \iota \, \frac{d \phi}{d t} = \pm \, j \, \Omega^\prime.
\end{equation} 
which is the same as the oscillation of QNMs.

To calculate $ \Omega^\prime $, firstly, we obtain the radius of unstable null circular equatorial geodesic orbits, $ r_0 $, from the following relations 
\begin{equation}\label{54_label}
	g_{t t} + 2 \, g_{t \phi} \, \frac{d \phi}{d t} + g_{\phi \phi} \, \big( \frac{d \phi}{d t} \big)^2 = 0,
\end{equation}
and
\begin{equation}\label{55_label}
	\Gamma_{t t}^r + 2 \, \Gamma_{t \phi}^r \, \frac{d \phi}{d t} + \Gamma_{\phi \phi}^r  \, \big( \frac{d \phi}{d t} \big)^2 = 0.
\end{equation}
where the former holds for every null path and the later is the radial component of the geodesic equation. We use  metric in Eq. \eqref{50_label} and simplify Eqs. \eqref{54_label} and \eqref{55_label} as follows
\begin{equation}\label{56_label}
	\begin{aligned}
		\frac{d \phi}{d t} & = \frac{a^2}{2 \, r^2 \, \sqrt{1 - 2 \, m/r}} \, \big[ 1 + 8 \, ( \frac{m}{r} )^2 \big]\\
		& + \sqrt{1 - \frac{2 \, m}{r}} \, \big[ 1 - q \, \ln(1 - \frac{2 \, m}{r})\big] + \frac{2 \, m \, a}{r^2},
	\end{aligned}
\end{equation}
\begin{equation}\label{57_label}
	\begin{aligned}
		\frac{d \phi}{d t} & = 1 - a \, \sqrt{\frac{m}{r^3}} + m \, \big( \frac{a}{r} \big)^2 - \frac{ \ln(1 - 2 \, m/r)}{2 (1 - 2 \, m/r)}\\
		& \times \big( \frac{4 \, m}{r} - \frac{q}{1 - 2 \, m/r} \big) + \frac{1}{2} \, \big( 1 - \frac{m}{r} \big).
	\end{aligned}
\end{equation}
By equating the right hand sides of Eqs. \eqref{56_label} and \eqref{57_label} and using the following perturbation
\begin{equation}\label{58_label}
	r = 3 \, m + \bar{\epsilon}.
\end{equation}
where $ 3 m $ is light ring radius of Schwarzschild black hole and $ \bar{\epsilon} $ is a perturbation parameter (and $r \equiv r_{0_{JNW}}$). Now, keep perturbation terms up to first order of $ \bar{\epsilon} $, we have 
\begin{equation}\label{59_label}
	r_{0_{JNW}}= 3 \, m \mp \frac{2 \, a}{\sqrt{3}} - \frac{2 \, a^2}{9 \, m} + 2 \, m \, q.
\end{equation}
By putting $ r_{0_{JNW}} $ in metric elements in Eq. \eqref{54_label} or Eq. \eqref{55_label}, we get
\begin{equation}\label{60_label}
	\Omega_\pm = \frac{d \phi}{d t} = \frac{1}{3 \, \sqrt{3} \, m} \pm \frac{1}{3 \, \sqrt{3} \, m} \, \big[ \frac{11}{54} \, ( \frac{a}{m} )^2 + \frac{2 \, \sqrt{3} \, a}{9 \, m} - q \, \ln 3 \big].
\end{equation} 
The other part of QNMs, $ \Gamma $, shows the decay rate of their amplitude. On the other hand, this rate corresponds to the divergence of the null rays on the circular orbit in the eikonal limit. To obtain $ \Gamma $, we should perturb the null equatorial circular orbit and therefore, we add a small perturbation in the coordinates $ \bar{x}^\mu = (t, r, \theta, \phi) = (t, r_0, \frac{\pi}{2}, \Omega^\prime t) $ as follows
\begin{equation}\label{61_label}
	r = r_0 \, [1 + \epsilon \, f(t)], \quad \phi = \Omega_\pm \, [t + \epsilon \, g(t)], \quad \ell = t + \epsilon \, h(t),
\end{equation}
where, $ \epsilon $ is a very small perturbation parameter and the initial conditions for $ f(t) $, $ g(t) $ and $ h(t) $ are as follows
\begin{equation}\label{62_label}
	f(0) = g(0) = h(0) = 0.
\end{equation}
It is clear from our definition of $ \Omega $ in Eq. \eqref{53_label} that $ g(t) = 0 $. 

The perturbed propagation vector (up t0 $ O(e^2) $ )  can be obtained as follows 
\begin{equation}\label{63_label}
	K^\mu = \frac{d x^\mu}{d \ell} = \big(1 - \epsilon \, h^\prime, \, \epsilon \, r_0 \, f^\prime, \, 0, \, \Omega_\pm \, (1 - \epsilon \, h^\prime)\big).
\end{equation}
where, prime refers to derivative with respect to t. If we assume $ \rho $ is the density of null rays the conservation law for a congruence of null rays  may be written as 
\begin{equation}\label{64_label}
	\begin{aligned}
		& \nabla_\mu \, (\rho \, K^\mu) = 0,\\
		& \Rightarrow \quad \frac{1}{\rho} \, \frac{d \rho}{d \ell} = - \nabla_\mu K^\mu = - \frac{1}{\sqrt{- g}} \, \frac{\partial}{\partial x^\alpha} \, (\sqrt{- g} \, K^\alpha).
	\end{aligned}
\end{equation}
If we neglect terms proportional to powers of $ \epsilon $, we have 
\begin{equation}\label{65_label}
	\frac{1}{\rho} \, \frac{d \rho}{d t} = - \frac{f^{\prime \prime}(t)}{f^\prime(t)}.
\end{equation} 
So, by having $ f(t) $, one can calculate $ \rho $. To obtain $ f(t) $, one can use the radial component of geodesic equation that is as follows
\begin{equation}\label{66_label}
	\frac{d^2 r}{d \ell^2} + \Gamma_{t t}^r  \,\big( \frac{d t}{d \ell} \big)^2 + \Gamma_{\phi \phi}^r \, \big( \frac{d \phi}{d \ell} \big)^2 + \Gamma_{\theta \theta}^r \, \big( \frac{d \theta}{d \ell} \big)^2 + 2 \, \Gamma_{t \phi}^r \, \frac{d t}{d \ell} \, \frac{d \phi}{d \ell} = 0.
\end{equation}
Now, if we put $ \theta = \pi/2 $ and expand Eq. \eqref{66_label} to the first order of $ \epsilon $ we have the following  differential equation for $ f(t) $
\begin{equation}\label{67_label}
	\begin{aligned}
		& 9 \, m^2 \, \big[ - 2 \, a \, (3 \, \sqrt{3} \, m + a) + 9 \, m^2 \, (3 + 2 \, q) \big] \, f^{\prime \prime}(t)\\
		& + \big[ 2 \, a \, (a + \sqrt{3} \, m) + 3 \, m^2 \, (2 \, q \, (3 \, \ln 3 - 1) - 3) \big] \, f(t) = 0.
	\end{aligned}
\end{equation} 
Therefore, $ f(t) $ is given by
\begin{equation}\label{68_label}
	f(t) = \sinh (\zeta t),
\end{equation}
where $ \zeta $ is
\begin{equation}\label{69_label}
	\zeta = \frac{1}{3 \, \sqrt{3} \, m} \, \big( 1 - \frac{2 \, a^2}{27 \, m^2} - q \, \ln 3  \big).
\end{equation}
By replacing $ f(t) $ in Eq. \eqref{65_label}, we derive the density of null rays as follows
\begin{equation}\label{70_label}
	\rho(t) = \rho(0) \, \frac{1}{\cosh(\zeta)} \simeq 2 \, \rho(0) \, (e^{- \zeta \, t} - e^{- 3 \, \zeta \, t} + e^{- 5 \, \zeta \, t} - \dots).
\end{equation}
Therefore one can derive the imaginary parts of the QNMs as follows 
\begin{equation}\label{71_label}
	\Gamma = \big( n + \frac{1}{2} \big) \, \zeta.
\end{equation}
Eventually, QNMs for the rotating JNW metric are given by
\begin{equation}\label{72_label}
	\begin{aligned}
		Q_{JNW} & = (\Omega + i \, \Gamma)_{JNW}\\
		& = j \, \big[ \frac{1}{3 \, \sqrt{3} \, m} \pm \frac{1}{3 \, \sqrt{3} \, m} \, \big( \frac{11}{54} \, ( \frac{a}{m} )^2 + \frac{2 \, \sqrt{3} \, a}{9 \, m}\\
		& - q  \,\ln 3 \big) \big] + i \, \big( n + \frac{1}{2} \big) \, \Big\{ \frac{1}{3 \, \sqrt{3} \, m} \big[ 1 - \frac{2 \, a^2}{27 \, m^2}\\
		& - q \, \ln 3 \big] \Big\}.
	\end{aligned}
\end{equation}
Eq. \eqref{72_label} in the limit $ a = 0 $ gives us the QNMs corresponding to the static JNW metric and in the limit $ q = 0 $ it gives the QNMs of the Kerr metric
\cite{ferrari1984new}.
\begin{figure}[t]
\centering
\includegraphics[width=1.0\columnwidth, trim={0.2cm 0.0cm 0.2cm 0.0cm}, clip]{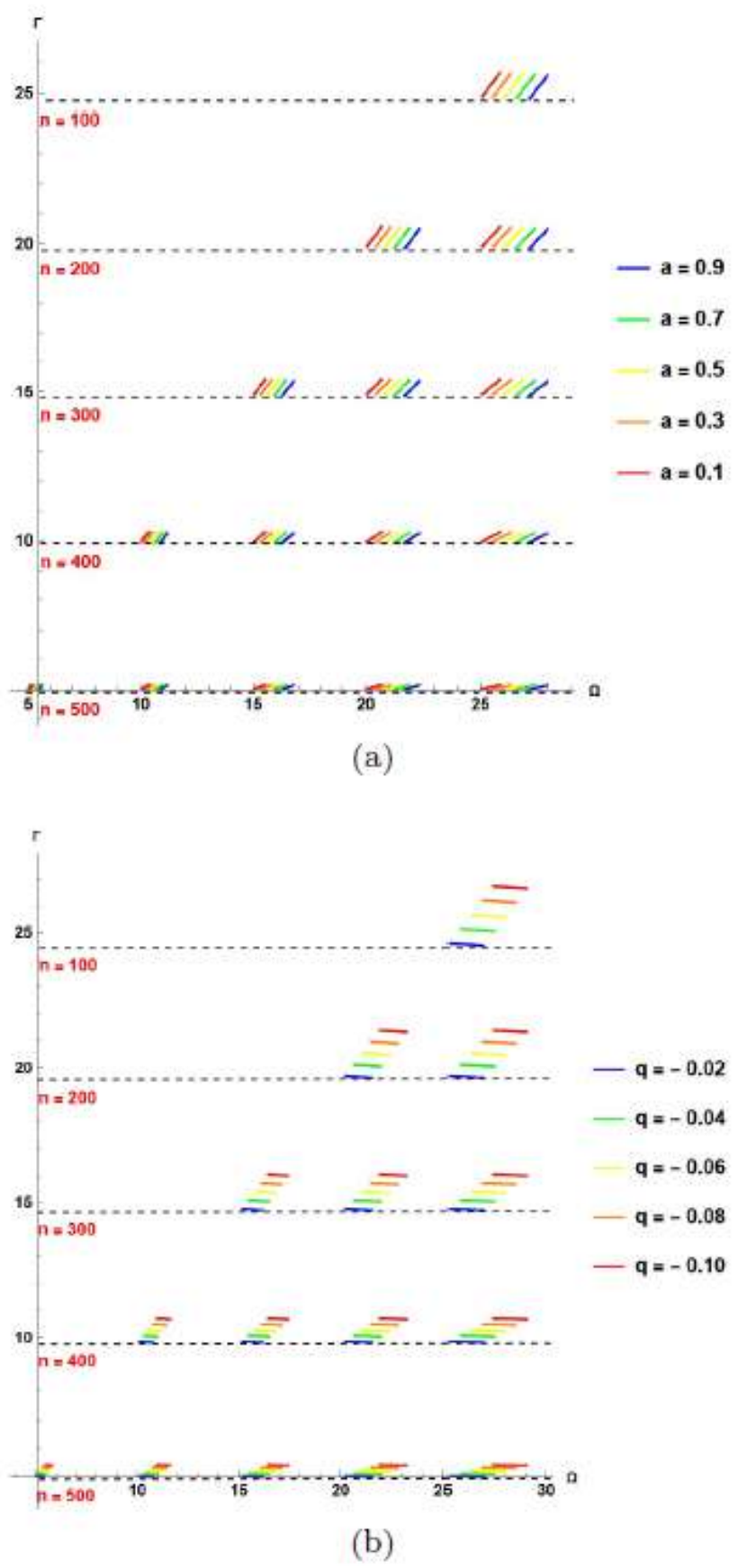}
\caption{$\Gamma-\Omega$ plot related to Eq. \eqref{72_label}, where $ m = 4 $, $n \in \{100, 200, 300, 400, 500\} $ and $ j \in \{ n, n + 100, \dots, 500 \} $. Dashed lines on the plots show $ n = 100 $ to $ n = 500 $ respectively from top to bottom. Panel (a): $ a = 0.1 $ (red), $ a = 0.3 $ (orange), $ a = 0.5 $ (yellow), $ a = 0.7 $ (green), $ a = 0.9 $ (blue) and $ q \in [-0.06,-0.03] $ . Panel (b): $ q = - 0.1 $ (red), $ q = - 0.08 $ (orange), $ q = - 0.06 $ (yellow), $ q = -0.04 $ (green), $ q = - 0.02 $ (blue) and $ a \in [0.3,0.9] $.}
\label{3fig}
\end{figure}

In Fig. \ref{3fig}, we have depicted the QNMs in Eq. \eqref{72_label}.
\subsection{QNMs of the BG metric}\label{63}
\begin{figure}[t]
\centering
\includegraphics[width=1.0\columnwidth, trim={0.2cm 0.0cm 0.2cm 0.0cm}, clip]{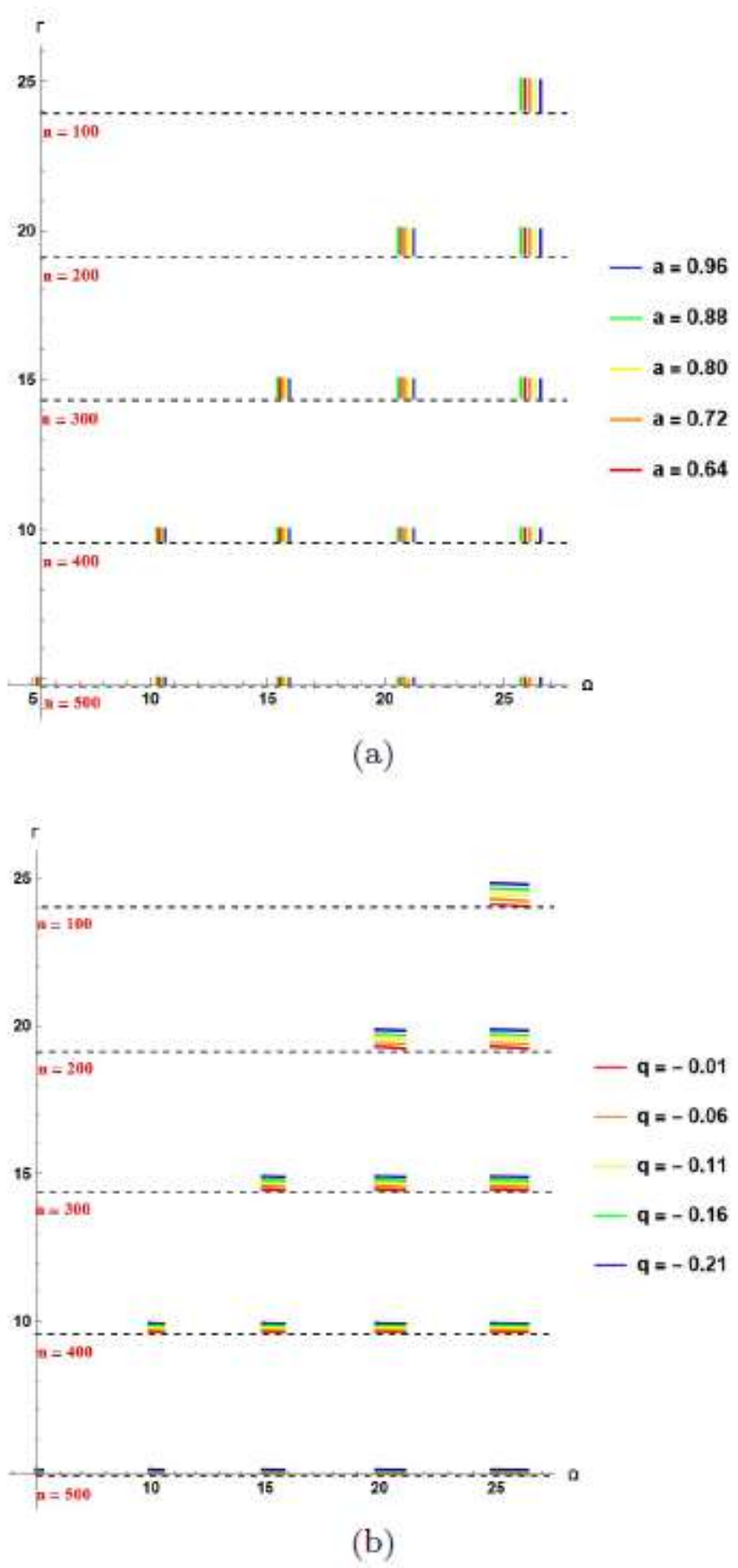}
\caption{$\Gamma-\Omega$ plot related to Eq. \eqref{74_label}, where $ m = 4 $, $n \in \{100, 200, 300, 400, 500\} $ and $ j \in \{ n, n + 100, \dots, 500 \} $. Dashed lines on the plots show $ n = 100 $ to $ n = 500 $ respectively from top to bottom. Panel (a): $ a = 0.64 $ (red), $ a = 0.72 $ (orange), $ a = 0.80 $ (yellow), $ a = 0.88 $ (green), $ a = 0.96 $ (blue) and $ \nu \in [-0.3,0) $ . Panel (b): $ \nu = - 0.01 $ (red), $ \nu = - 0.06 $ (orange), $ \nu = - 0.11 $ (yellow), $ \nu = -0.16 $ (green), $ \nu = - 0.21 $ (blue) and $ a \in [0.3,0.9] $.}
\label{4fig}
\end{figure}

In this part, we are going to calculate the QNMs of the BG metric by using the light-ring method. 
By replacing transformations \eqref{14_label} in the metric \eqref{8_label} and with condition $ q = 0 $ in Eqs. (\ref{11_label}-\ref{13_label}) and expanding the results to first order in $ \nu $ and second order in $ a $ and neglecting terms proportional to $ a \nu $, we have
\begin{equation}\label{73_label}
\begin{aligned}
d s^2 = & - \big\{ 1 - \frac{2 \, m}{r} + \frac{2 \, m \, a^2}{r^3} \cos^2 \theta \big\} \, d t^2\\
& + \frac{1}{1 - 2 \, m/r} \, \Big\{ 1 - \frac{a^2}{r^2 \, (1 - 2 \, m/r)}\\
&\times \big[ \cos^2 \theta \, (\frac{2 \, m}{r} - 1) + 1 \big] + \nu \, \big[ \ln \big((1 - \frac{m}{r})^2\\
& - \frac{m^2}{r^2} \cos^2 \theta\big) - \ln\small(1 - \frac{2\,m}{r}\small) \big] \Big\} \, d r^2\\
& + \Big\{ r^2 + a^2 \cos^2 \theta  + \nu \, r^2 \,\big[ \ln \big((1 - \frac{m}{r})^2\\
& - \frac{m^2}{r^2} \cos^2 \theta\big) - \ln \small(1 - \frac{2 \, m}{r}\small) \big] \Big\} \, d \theta^2\\
& + \frac{\sin^2\theta}{r} \big\{ r^3+a^2\,(r+m)-a^2\,m\,\cos(2\theta) \big\} \, d \phi^2\\
&+ \frac{4 \, a \, m}{r} \sin^2 \theta \,  d t \, d \phi.
\end{aligned}
\end{equation}
By performing calculations similar to subsection \ref{62}, QNMs to the BG metric are obtained as follows
\begin{equation}\label{74_label}
	\begin{aligned}
		Q_{BG} & = (\Omega + i \Gamma)_ {BG}\\
		& = j \, \big[ \frac{1}{3 \, \sqrt{3} \, m} \pm \frac{1}{3 \, \sqrt{3} \, m} \big( \frac{11}{54} \, (\frac{a}{m})^2 + \frac{2\,\sqrt{3}\,a}{9\,m} \big) \big]\\
		& + i \, \big( n + \frac{1}{2} \big) \, \Big\{ \frac{1}{3 \, \sqrt{3} \, m} \big[ 1 - \frac{2 \, a^2}{27 \, m^2} - \nu \, (\ln 2\\
		& - \frac{\ln 3}{2} ) \big] \Big\}.
	\end{aligned}
\end{equation}
Fig. \ref{4fig} shows the QNMs related to the BG metric.
\subsection{QNMs for a rotating metric with $ \mu=0 \; (\nu=1-\gamma) $}\label{7}
Now, we intend to obtain the QNMs corresponding to the case where $ \mu = 0 $ ($ \nu = 1 - \gamma $) in Eq. \eqref{11_label}. By replacing transformations \eqref{14_label} in the metric \eqref{8_label} and expanding the result to first order in perturbation parameter $ q $ and second order in rotation parameter $ a $ and neglecting terms proportional to $ a q $, we have
\begin{equation}\label{75_label}
	\begin{aligned}
		d s^2 & = - \, \big\{ 1 - \frac{2 \, m}{r} + \frac{2 \, m \, a^2}{r^3} \cos^2 \theta  + q \, (1 - \frac{2 \, m}{r})\\
		&\times \ln \small(1 - \frac{2 \, m}{r}\small) \big\} d t^2 + \frac{1}{1 - 2 \, m/r} \, \Big\{ 1 - \frac{a^2}{r^2 \, (1 - 2 \, m/r)}\\
		&\times [\cos^2 \theta \, (\frac{2 \, m}{r} - 1) + 1]  - q \, \ln \big[(1 - \frac{m}{r})^2\\
		& - \frac{m^2}{r^2} \cos^2 \theta \big] \Big\} \, d r^2\\
		& + \Big\{ r^2 + a^2 \, \cos^2 \theta  - q \, r^2 \, \ln \big[ (1 - \frac{m}{r})^2\\
		& - \frac{m^2}{r^2} \cos^2 \theta \big] \Big\} \, d \theta^2\\
		& + \sin^2 \theta \, \big\{ \frac{1}{r}\big[r^3+a^2\,\big(m+r-m\,\cos(2\,\theta)\big)\big]\\
		& - q \, r^2 \, \ln \small(1 - \frac{2 \, m}{r}\small) \big\} \, d \phi^2\\
		&+ \frac{4 \, a \, m}{r} \sin^2 \theta \, d \phi \, d t.
	\end{aligned}
\end{equation}
Similar to the previous subsection, the QNMs of this metric are obtained as follows $ (\nu = - q) $
\begin{equation}\label{76_label}
	\begin{aligned}
		Q_{\nu=-q} & = (\Omega + i \, \Gamma)_{\nu=-q}\\
		& = j \, \big[ \frac{1}{3 \, \sqrt{3} \, m} \pm \frac{1}{3 \, \sqrt{3} \, m} \big( \frac{11}{54} \, ( \frac{a}{m} )^2 + \frac{2 \, \sqrt{3} \, a}{9 \, m}\\
		& - q \, \ln 3 \big) \big] + i \, \big( n + \frac{1}{2} \big) \, \Big[ \frac{1}{3 \, \sqrt{3} \, m} \big( 1 - \frac{2 \, a^2}{27 \, m^2}\\
		& + q \, ( \ln 2 - \frac{3\,\ln3}{2}) \big) \Big].
	\end{aligned}
\end{equation}
According to Eqs. \eqref{72_label}, \eqref{74_label} and \eqref{76_label}, it is clear that in the real part of the QNMs ($\Omega$), the first three terms are equal for all metrics, and the rotating JNW metric and the rotating metric at $ \nu=-q $ have an additional term $ (q\,\ln 3) $ compared to the BG metric. In the BG metric, since $q = 0$, $ (q\,\ln 3) $, term does not exist.

Considering the equality of the real parts of the rotating JNW metric $ (\nu=0) $ and the rotating metric at $ \nu=-q $, it is clear that the frequencies emitted from the rotating metrics are the same.

In the imaginary part of QNMs, the first two terms are the same for all three metrics and the rest of the terms for rotating JNW and BG metrics and rotating metric at $ \nu=-q $ are equal to $ - 1.10\,q $, $ -0.14\,\nu $ and $ -0.95\,q $ respectively. 
\section{Precession frequency of rotating metrics}\label{6100}
General relativity states that the spin of a gyroscope precess for two reasons, one is due to the curvature of space-time caused by the existence of the object and the other is due to the rotation of the object.

The spin precession of a gyroscope according to the presence of matter was first described by Willem de Sitter in 1916 and is known as the de Sitter effect (also called  geodetic precession or de Sitter precession)
\cite{de1916einstein}.

When the object rotates, the inertial frame is dragged along it. Also, the rotation of a massive object causes the precession of the spin in a test gyroscope close to it. This effects was first stated by Josef Lense and Hans Thirring (1918) and is known as Lense-Thirring (LT) precession or LT effect
\cite{lense1918influence}.

In this section, we want to investigate the LT and de Sitter effects on the metrics of the previous sections.

The precession frequency vector  in Copernican frame for any arbitrary stationary metric  is as follows 
\cite{straumann2013applications}
\begin{equation}\label{80_label}
	\Omega_{LT} = \frac{1}{2} \, \frac{\epsilon_{i j k}}{\sqrt{- g}} \, \big[ \partial_j \, g_{0 i} \, \big( \partial_k - \frac{g_{0 k}}{g_{0 0}} \, \partial_0 \big) - \frac{g_{0 i}}{g_{0 0}} \, \partial_j \, g_{0 0} \, \partial_k \big].
\end{equation}
where $ \epsilon_{i j k} $ is the Levi-Civita symbol. In the case of axially symmetric metrics that we are dealing with, Eq. \eqref{80_label} becomes the following relation
\begin{equation}\label{81_label}
	\begin{aligned}
		\Omega_{LT} & = \Omega^r \, \partial_r +  \Omega^\theta \, \partial_\theta = \frac{1}{2 \, \sqrt{- g}} \big[ - \big( \partial_\theta \, g_{t \phi} - \frac{g_{t \phi}}{g_{t t}} \, \partial_\theta \, g_{t t} \big) \, \partial_r\\
		& +\big( \partial_r \, g_{t \phi} - \frac{g_{t \phi}}{g_{t t}} \, \partial_r \, g_{t t} \big) \, \partial_\theta \big].
	\end{aligned}
\end{equation}
The magnitude of the vector in Eq. \eqref{81_label} is
\begin{equation}\label{83_label}
	\Omega_{LT}  = \sqrt{g_{rr} \, \Omega_r^2 + g_{\theta \theta} \, \Omega_\theta^2}.
\end{equation}
Now, in addition to the LT precession effect, if we also consider the de Sitter effect, the relation of the total precession frequency vector is as follows
\cite{chakraborty2017distinguishing}
\begin{equation}\label{86_label}
	\begin{aligned}
		\Omega_F & = \frac{\epsilon_{i j k}}{2 \, \sqrt{- g} \, \big( 1 + 2 \, \Omega \, \frac{g_{0 i}}{g_{0 0}} + \Omega^2 \, \frac{g_{i i}}{g_{0 0}} \big)}\\
		& \big[ \big( \partial_j \, g_{0 i} - \frac{g_{0 i}}{g_{0 0}} \, \partial_j \, g_{0 0} \big) + \Omega \, \big( \partial_j \, g_{i i} - \frac{g_{i i}}{g_{0 0}} \, \partial_j \, g_{0 0} \big)\\
		& + \Omega^2 \, \big(  \frac{g_{0 i}}{g_{0 0}} \, \partial_j \, g_{i i} - \frac{g_{i i}}{g_{0 0}} \, \partial_j \, g_{0 i} \big) \big] \, \partial_k.
	\end{aligned}
\end{equation}
where $ \Omega $ represents the angular velocity of the gyroscope.

Eq. \eqref{86_label} for axisymmetric space time becomes
\begin{equation}\label{87_label}
	\begin{aligned}
		& \Omega_F = \frac{1}{2 \, \sqrt{- g} \, \big( 1 + 2 \, \Omega \, \frac{g_{t \phi}}{g_{t t}} + \Omega^2 \, \frac{g_{\phi \phi}}{g_{t t}} \big)}\\
		& \big[ - \big[ \big( \partial_\theta \, g_{t \phi} - \frac{g_{t \phi}}{g_{t t}} \, \partial_\theta \, g_{t t} \big) + \Omega \, \big( \partial_\theta \, g_{\phi \phi} - \frac{g_{\phi \phi}}{g_{t t}} \, \partial_\theta \, g_{t t} \big)\\
		& + \Omega^2 \, \big(  \frac{g_{t \phi}}{g_{t t}} \, \partial_\theta \, g_{\phi \phi} - \frac{g_{\phi \phi}}{g_{t t}} \, \partial_\theta \, g_{t \phi} \big) \big] \, \partial_r\\
		& + \big[ \big( \partial_r \, g_{t \phi} - \frac{g_{t \phi}}{g_{t t}} \, \partial_r \, g_{t t} \big) + \Omega \, \big( \partial_r \, g_{\phi \phi} - \frac{g_{\phi \phi}}{g_{t t}} \, \partial_r \, g_{t t} \big)\\
		& + \Omega^2 \, \big(  \frac{g_{t \phi}}{g_{t t}} \, \partial_r \, g_{\phi \phi} - \frac{g_{\phi \phi}}{g_{t t}} \, \partial_r \, g_{t \phi} \big) \big] \, \partial_\theta \big].
	\end{aligned}
\end{equation}
The magnitude of total  precession frequency vector is
\begin{equation}\label{102_label}
	\Omega_F  = \sqrt{g_{rr} \, \Omega_{F_r}^2 + g_{\theta \theta} \, \Omega_{F_\theta}^2}.
\end{equation}
In the following  we derive precession frequencies of the rotating metrics that were introduced in Sec. \ref{5}. Also, we use the approximations from the previous section to calculate the precession frequencies.
\subsection{Precession frequency in rotating JNW metric}
In this part, we are going to calculate the precession frequency for rotating JNW metric. For this, we use the metric in Eq. \eqref{50_label}. 

Using Eqs. \eqref{81_label} and \eqref{83_label} for the rotating JNW metric, we have
\begin{equation}\label{88_label}
	\Omega_{LT}^{JNW}  =  \frac{a \, m}{r^2\,(r-2\,m)}.
\end{equation}
The above relation shows the precession frequency due to the LT effect for the rotating JNW metric. Now we are going to obtain the total precession frequency for the rotating JNW metric by combining the de Sitter and LT effects. Therefore, by using Eqs. \eqref{87_label} and \eqref{102_label} and expanding the result to first order in $ a $ and $ m $ and negative third order $ r $, we have
\begin{equation}\label{89_label}
	\Omega_F^{JNW}  =  \frac{1 + r \, \Omega^2 \, \big\{ r - m \, [3 \, (1 - a \, \Omega) + 4 \, q] \big\}}{r^3 \, \Omega^3}.
\end{equation}
\subsection{Precession frequency for the BG metric}
By using the metric in Eq. \eqref{73_label} and Eqs. \eqref{81_label} and \eqref{83_label}, we have
\begin{equation}\label{90_label}
	\Omega_{LT}^{BG}  = \Omega_{LT}^{JNW}.
\end{equation}
To obtain the total precession frequency for BG metric, we use Eqs. \eqref{87_label} and \eqref{102_label}  and expanding the result to first orders in $ a $, $ m $ and $ r^{-3} $ 
\begin{equation}\label{91_label}
	\Omega_F^{BG}  =  \frac{1 + r \, \Omega^2 \, \big\{ r - m \, [3 \, (1 - a \, \Omega) ] \big\} }{r^3 \, \Omega^3}.
\end{equation}
\subsection{Precession frequency for a rotating metric with $ \mu=0 \; (\nu = 1 - \gamma $)}
According to metric \eqref{75_label} and Eqs. \eqref{81_label} and \eqref{83_label}, $ \Omega_{LT} $ in this case $ (\mu=0) $ is equal to
\begin{equation}\label{92_label}
	\Omega_{LT}^{\nu=-q}  = \Omega_{LT}^{JNW}.
\end{equation}
The total Precession frequency $ \Omega_F $ for this case is equal to the total Precession frequency of the rotating JNW metric, i.e. Eq. \eqref{89_label}.
\begin{figure}[t]
\centering
\includegraphics[width=1.0\columnwidth, trim={0.2cm 0.0cm 0.2cm 0.0cm}, clip]{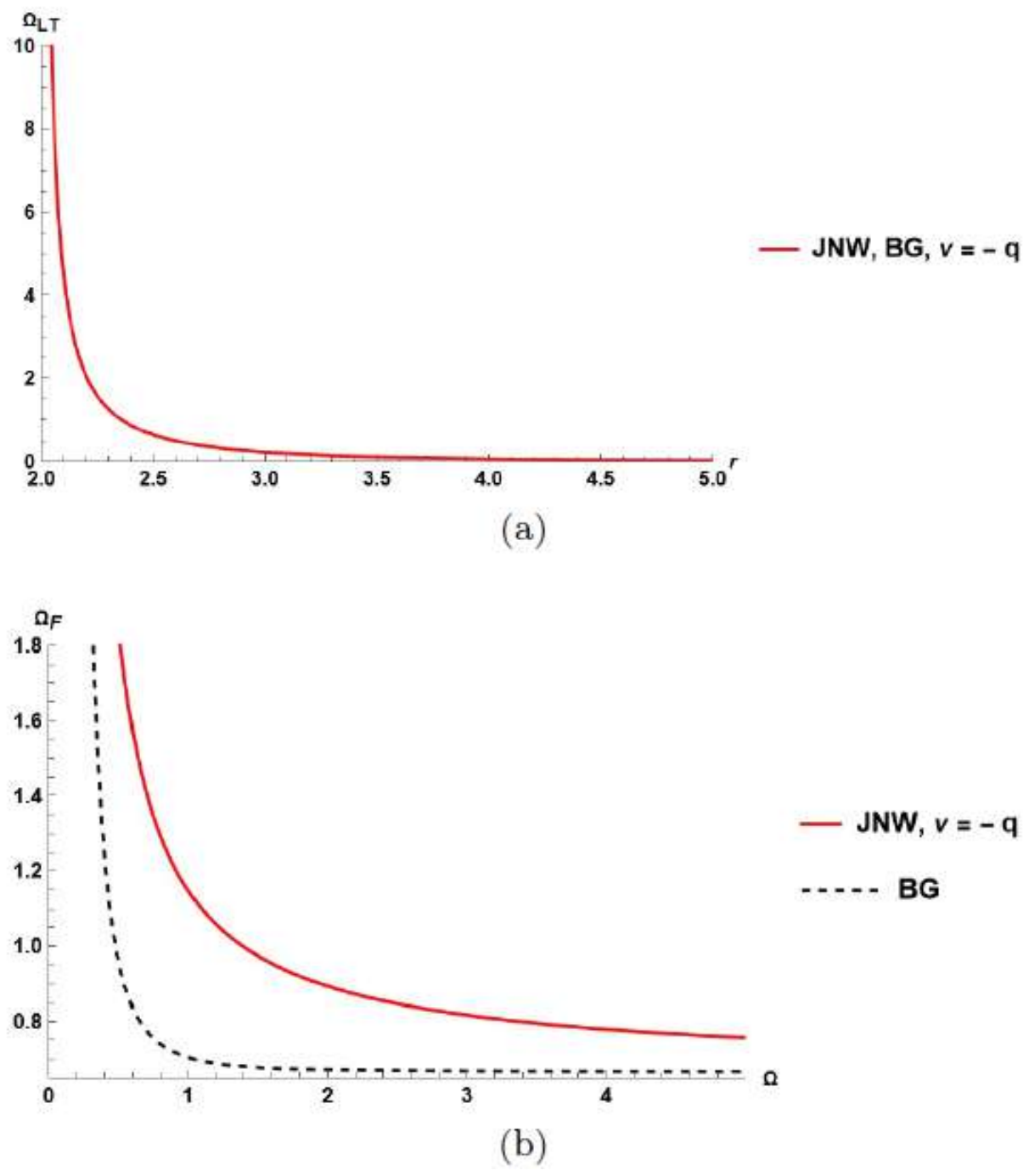}
\caption{The behavior of Precession frequency for $ m = 1 $, $ a = 2 $ and $ q = - 1 $. Panel (a): $ \Omega_{LT} $ vs $ r $ for rotating JNW and BG metrics and rotating metric at $ \nu=-q $ (solid red). According to Eqs. \eqref{88_label}, \eqref{90_label} and \eqref{92_label} $ \Omega_{LT} $ is infinite in $ r = 2\,m $. Panel (b): $ \Omega_{F} $ vs $ \Omega $ with $ r = 3 $ for rotating JNW metric and rotating metric  at $ \nu = -q $ (solid red) and BG metric (black dash). By increasing the angular velocity of the gyroscope, the related precession frequency decreases.}
\label{5fig}
\end{figure}

Fig. \ref{5fig} is related to the precession frequency. Panel a shows the graph of $ \Omega_{LT} $ versus $ a $, and the behavior of $ \Omega_{F} $ versus $ \Omega $ is depicted in panel b.

The Eqs. \eqref{88_label}, \eqref{90_label} and \eqref{92_label} show that the precession frequency due to the LT effect is the same for rotating JNW and BG metrics and rotating metric at $ \nu=-q $. Considering the total precession frequency ($\Omega_{F}$), both the rotating JNW metric and the rotating metric at $ \nu = -q $ have an additional term compared to the BG metric ($q=0$). The total frequency of the BG metric is greater than in the other cases.

By comparing the QNMs with the precession frequencies, we conclude that the method of QNMs is more accurate for observing the differences between the investigated rotating metrics.
\section{Conclusion}\label{7100} 
We considered a class of three parameter static metrics that becomes $ \gamma$-metric and JNW metric in certain values of the parameters
\cite{Azizallahi2023}.
We found a general class of rotating forms of the three parameter metrics. The rotating solutions in certain values of the parameters turns into the well known rotating solution of the $ \gamma$-metric. Also, we can find Bogush-Gal’tsov (BG) metric.

We obtained a correct form of the rotating JNW metric and we studied its Kretschmann scalar, quasi normal modes (QNMs) and precession frequency. Many different new calculations can be done on this large class of  rotating metrics in the future. For example, one may consider exact and non-perturbative forms of the rotating metrics and obtain QNMs and precession frequencies for them by numerical methods.
\section*{Appendix}
the equations of motion using Eq. \eqref{2_label} and metric in Eq. \eqref{500_label} are as follows
\begin{equation}\label{501_label}
	\begin{aligned}
		R_{t \rho}  &= R_{\rho t} = R_{t z} = R_{z t} = R_{\rho \phi} = R_{\phi \rho} = R_{z \phi} = R_{\phi z}\\
		& = R_{\phi \phi} = 0,
	\end{aligned}
\end{equation}
\begin{equation}\label{501a_label}
	\begin{aligned}
		R_{t t} & =  \frac{e^{- 2 \, \eta}}{2 \, \sigma^2} \, \Big\{ f \, (\partial_\rho^2 \, f + \partial_z^2 \, f) - (\partial_\rho \, f)^2 - (\partial_z \, f)^2\\
		& + \frac{f^4}{\rho^2 \sigma^2} \, [(\partial_\rho \, \omega)^2 + (\partial_z \, \omega)^2] + \frac{1}{\rho} \, f \, \partial_\rho \, f \Big\} = 0,
	\end{aligned}
\end{equation}
\begin{equation}\label{501b_label}
	\begin{aligned}
		R_{t \phi} & = R_{\phi t} = \frac{e^{- 2 \, \eta}}{2 \, \sigma^2} \, \Big\{ \omega \, \big[(\partial_\rho \, f)^2 + (\partial_z \, f)^2 - f \, (\partial_\rho^2 \, f\\
		& + \partial_z^2 \, f )\big] - f^2 \, (\partial_\rho^2 \, \omega+ \partial_z^2 \, \omega) - 2 \, f \, (\partial_\rho \, f  \, \partial_\rho \, \omega\\
		& + \partial_z \, f \, \partial_z \, \omega)  - \frac{f^4 \, \omega}{\rho^2 \, \sigma^2} \, \big[ (\partial_\rho \, \omega)^2 + (\partial_z \, \omega)^2 \big]\\
		& + \frac{f}{\rho} \, (f \, \partial_\rho \, \omega - \omega \, \partial_\rho \, f) \Big\} = 0,
	\end{aligned}
\end{equation}
\begin{equation}\label{501c_label}
	\begin{aligned}
		R_{\rho \rho} & = - \partial_\rho^2 \, \eta - \partial_z^2 \, \eta + \frac{1}{2 \, f} \, \Big\{ \partial_\rho^2 \, f + \partial_z^2 \, f + \frac{1}{\rho} \, \partial_\rho \, f\\
		& - \frac{1}{f} \, \big[ 2 \, (\partial_\rho \, f)^2 + (\partial_z \, f)^2\big] \Big\} + \frac{1}{\rho} \, \big(\partial_\rho \, \eta + \frac{f^2 \, (\partial_\rho \, \omega)^2}{2 \, \rho \, \sigma^2}\big)\\
		& = 8 \, \pi \, \partial_\rho \, \varphi \, \partial_\rho \, \varphi,
	\end{aligned}
\end{equation}
\begin{equation}\label{501d_label}
	\begin{aligned}
		R_{\rho z} & = R_{z \rho} = \frac{1}{\rho} \, \big(\partial_z \, \eta + \frac{f^2 \, \partial_\rho \, \omega \, \partial_z \, \omega}{2 \, \rho \, \sigma^2}\big) - \frac{\partial_\rho \, f \, \partial_z \, f}{2 \, f^2}\\
		& = 8 \, \pi \, \partial_\rho \, \varphi \, \partial_z \, \varphi,
	\end{aligned}
\end{equation}
\begin{equation}\label{501e_label}
	\begin{aligned}
		R_{z z} & = - \, \partial_\rho^2 \, \eta - \partial_z^2 \, \eta + \frac{1}{2 \, f} \, \Big\{ \partial_\rho^2 \, f + \partial_z^2 \, f + \frac{1}{\rho} \, \partial_\rho \, f\\
		& - \frac{1}{f} \, \big[ (\partial_\rho \, f)^2 + 2 \, (\partial_z \, f)^2 \big] \Big\} + \frac{1}{\rho} \, \big(- \partial_\rho \, \eta\\
		& + \frac{f^2 \, (\partial_z \, \omega)^2}{2 \, \rho \, \sigma^2} \big) = 8 \, \pi \, \partial_z \, \varphi \, \partial_z \, \varphi.
	\end{aligned}
\end{equation} 
\section*{Acknowledgements}
We would like to thank Yuri Pirogov and Oleg Zenin for useful comments. We acknowledge Isfahan University of Technology for the financial support that was made available to us.
	
	\newpage
	\bibliography{bibliography}{}
	
\end{document}